\documentclass[sigconf]{acmart}
\usepackage{multirow,tabularx,caption,subcaption,bm,enumitem,kantlipsum,xcolor,balance}

\AtBeginDocument{%
  \providecommand\BibTeX{{%
    \normalfont B\kern-0.5em{\scshape i\kern-0.25em b}\kern-0.8em\TeX}}}

\copyrightyear{2024}
\acmYear{2024}
\setcopyright{rightsretained}
\acmConference[WSDM '24] {Proceedings of the 17th ACM International Conference on Web Search and Data Mining}{March 4--8, 2024}{Merida, Mexico.}
\acmBooktitle{Proceedings of the 17th ACM International Conference on Web Search and Data Mining (WSDM '24), March 4--8, 2024, Merida, Mexico}
\acmPrice{15.00}
\acmISBN{979-8-4007-0371-3/24/03}
\acmDOI{10.1145/3616855.3635805}

\makeatletter
\gdef\@copyrightpermission{
  \begin{minipage}{0.3\columnwidth}
  \href{https://creativecommons.org/licenses/by/4.0/}{\includegraphics[width=0.90\textwidth]{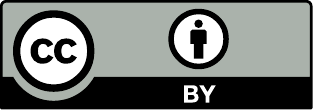}}  
\end{minipage}\hfill
  \begin{minipage}{0.7\columnwidth}
  \href{https://creativecommons.org/licenses/by/4.0/}{This work is licensed under a Creative Commons Attribution International 4.0 License.}
  \end{minipage}
  \vspace{5pt}
}
\makeatother
\begin{document}

\title[Exploring Adapter-based Transfer Learning
for Recommender Systems]{Exploring Adapter-based Transfer Learning
for Recommender Systems: Empirical Studies and Practical Insights 
}

\author{Junchen Fu} 
\affiliation{
  \institution{Westlake University}\streetaddress{}\city{}\country{}}
\email{ron.junchen.fu@gmail.com}
  
\author{Fajie Yuan$^{\dagger}$}
\affiliation{
\institution{Westlake University}\streetaddress{}\city{}\country{}}
\email{yuanfajie@westlake.edu.cn}

\author{Yu Song}
\affiliation{
\institution{Westlake University}\streetaddress{}\city{}\country{}}
\email{songyu@westlake.edu.cn}

\author{Zheng Yuan}
\affiliation{
\institution{Westlake University}\streetaddress{}\city{}\country{}}
\email{yuanzheng@westlake.edu.cn}

\author{Mingyue Cheng}
\affiliation{
\institution{University of Science and Technology of China}\streetaddress{}\city{}\country{}}
\email{mycheng@ustc.edu.cn}

\author{Shenghui Cheng}\affiliation{
\institution{Westlake University}\streetaddress{}\city{}\country{}}
\email{chengshenghui@westlake.edu.cn}

\author{Jiaqi Zhang}\affiliation{
\institution{Westlake University}\streetaddress{}\city{}\country{}}
\email{zhangjiaqi@westlake.edu.cn}

\author{Jie Wang$^{*}$}\affiliation{
\institution{Westlake University}\streetaddress{}\city{}\country{}}
\email{j.wang.9@research.gla.ac.uk}

\author{Yunzhu Pan}\affiliation{
\institution{Westlake University}\streetaddress{}\city{}\country{}}
\email{panyunzhu@westlake.edu.cn}

\thanks{$\dagger$ Corresponding author. Fajie designed and supervised this research; Junchen performed this research, in charge of key technical parts; Junchen, Fajie, and Yu wrote the manuscript. Other authors assisted in partial experiments. }
\thanks{* The work was done when Jie Wang was a visiting scholar at Westlake University. Her current affiliation is with the University of Glasgow.}

\renewcommand{\shortauthors}{Fu, et al.}

\begin{abstract}
Adapters, a plug-in neural network module with some tunable parameters, have emerged as a parameter-efficient transfer learning technique for adapting pre-trained models to downstream tasks, especially for natural language processing (NLP) and computer vision (CV) fields. Meanwhile, learning recommendation models directly from  raw item modality features ---  e.g., texts of NLP and images of CV --- can enable effective and \underline{trans}ferable \underline{rec}ommender systems (called TransRec). In view of this, a natural question arises: \textit{can  adapter-based learning techniques achieve parameter-efficient TransRec with good  performance?}

To this end, we perform empirical studies 
to address several key sub-questions.
First, we ask \textit{whether the adapter-based TransRec performs comparably to  TransRec based on standard full-parameter fine-tuning? does it hold for recommendation with different item modalities, e.g., textual RS and visual RS.} If yes, we benchmark these existing adapters, which have been shown to be effective in  NLP and CV tasks, in item recommendation tasks. Third, we carefully study several key factors for the adapter-based TransRec in terms of \textit{where and how to insert these adapters?} Finally, we look at the effects of adapter-based TransRec by either scaling up its source training data or scaling down its target training data. Our paper provides key insights and practical guidance on unified \& transferable recommendation --- a less studied recommendation scenario.
We release our codes and other materials at \color{blue}{\url{https://github.com/westlake-repl/Adapter4Rec/}}.

\end{abstract}



 \maketitle

\section{Introduction}

Recently, large foundation models~\cite{bommasani2021opportunities} have attracted considerable attention in the entire AI community. BERT~\cite{devlin2018bert}, GPT-3~\cite{brown2020language}, CLIP~\cite{radford2021learning} and various Vision Transformers (ViT)~\cite{dosovitskiy2020image,Liu_2021_ICCV} have demonstrated impressive transfer learning capabilities on a range of benchmark tasks, and are now reshaping the paradigm of the natural language processing (NLP) and computer vision (CV) communities. 
Inspired by the enormous success, the research of developing pre-trained \& \underline{trans}ferable \underline{rec}ommender system (TransRec)  models is becoming increasingly popular as well~\cite{yuan2020parameter,yuan2021one,shin2021one4all,zhang2023ninerec,cheng2023image,wang2023missrec,yang2023collaborative}. 
TransRec offers a natural solution to address the challenges of sparsity and insufficient data in recommender systems (RS) through the application of transfer learning. Specifically, TransRec leverages the knowledge acquired from larger data sources, including user/item representations and their corresponding matching relationships, and transfers this knowledge to the current RS, which might have limited data availability. 

\begin{figure}[h]
  \centering
  \includegraphics[width=\linewidth]{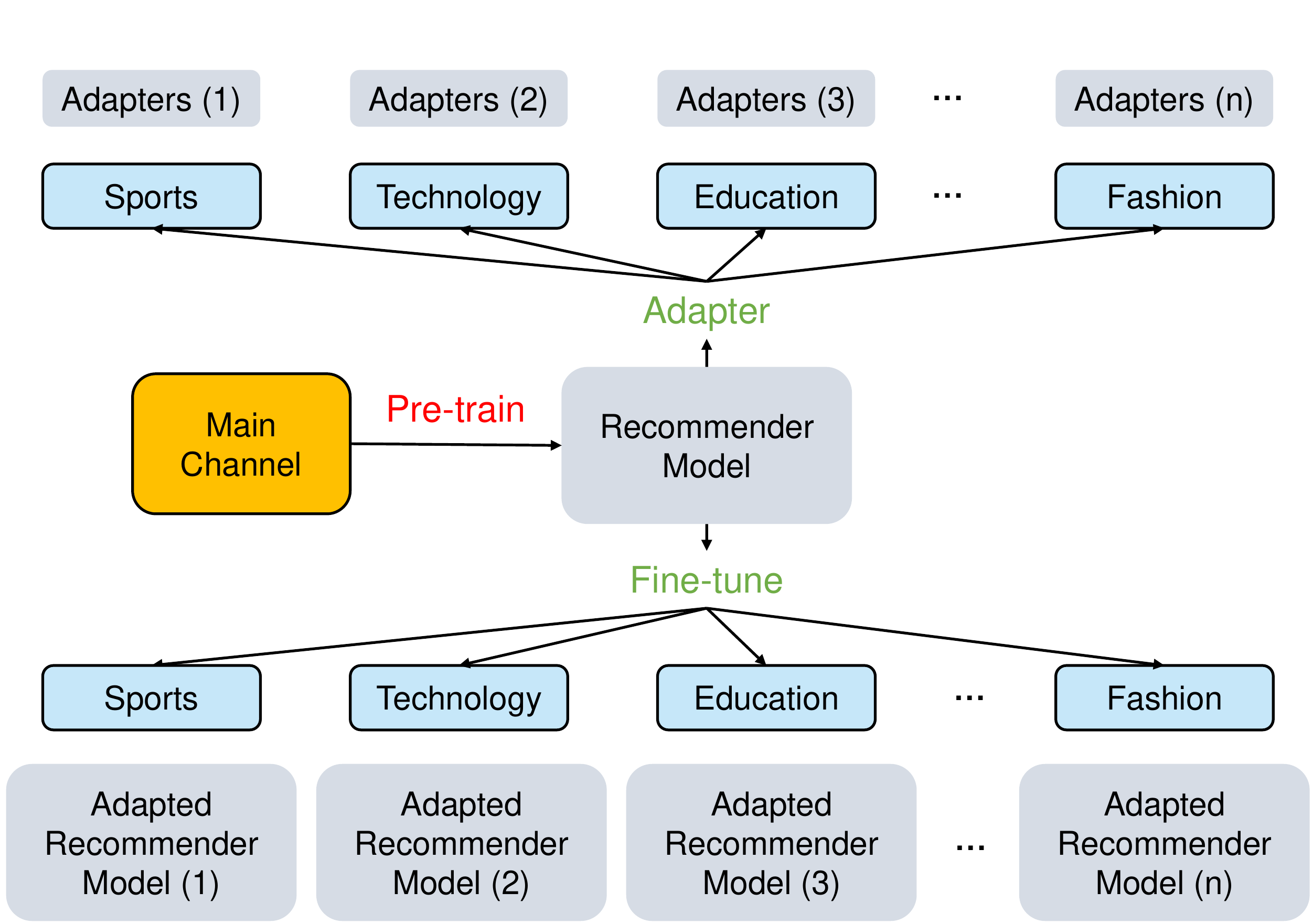}
  \caption{Large industrial RS platforms often have a main recommendation channel and various vertical channels, e.g., sports, education, fashion, etc.
  One has to maintain the entire model for each channel by a separate model with standard fine-tuning; by contrast, only a small set of parameters needs to be maintained by adapter tuning.  }
  \Description{}
  \label{fig:Multi_cate}
\end{figure}

In fact, prior to the era of large-scale foundation models, significant research efforts were dedicated to studying TransRec, particularly for cross-domain or cold-start recommendations~\cite{li2009transfer,pan2010transfer,pan2011transfer}. For example, PeterRec~\cite{yuan2020parameter}, ShopperBERT~\cite{shin2021one4all}, Conure~\cite{yuan2021one}, and STAR~\cite{sheng2021one},  applied modern deep neural networks to transfer user- or item-level preference across different recommendation platforms. However, these works are mainly ID-based collaborative filtering (IDRec), which highly relies on overlapped userID or itemID data when transferring knowledge. 
This strict overlapping assumption hardly holds in practice~\cite{yuan2023go,zhang2023ninerec} --- e.g., TikTok is unlikely to share their userIDs or itemIDs to YouTube, and vice versa.

To realize more general transfer learning,  the common practice is to represent items with their raw modality features (e.g., text or images) and users with the interacted item sequence\footnote{From this perspective, the transferable recommendation models in the existing literature are mostly a sequential model.
}
rather than userIDs and itemIDs \cite{ding2021zero, wang2022transrec,shin2021scaling,yuan2023go,cheng2023image,zhang2023ninerec,li2023exploring}. By replacing ID embeddings with powerful item modality encoders (ME), such as BERT and ViT, TransRec has shown state-of-the-art results on many downstream tasks. 

According to the above works, the modern TransRec framework typically consists of two modules, i.e.,  a user encoder with one or multiple item ME. TransRec models are usually initially pre-trained on extensive upstream recommendation data and subsequently fine-tuned to cater to various downstream recommendation tasks. 
Here, we argue that the commonly adopted full parameter
fine-tuning in TransRec has several key issues:

 \begin{figure}[h]
  \centering
  \includegraphics[width=0.9\linewidth]{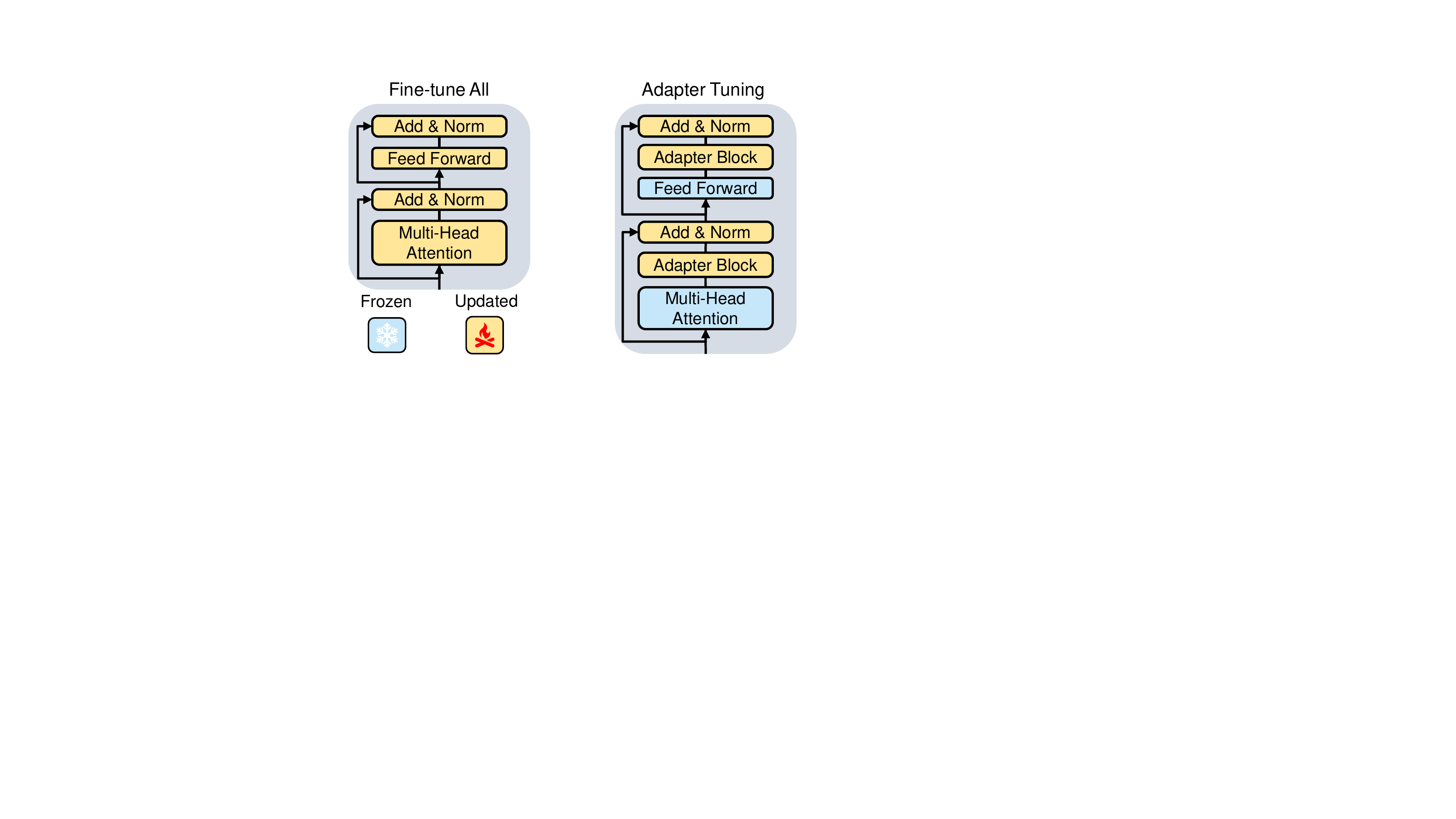}
  \caption{Fine-tune All (FTA) vs. Adapter Tuning (AdaT). 
  The FTA method updates all parameters in the Transformer~\cite{vaswani2017attention} block. On the contrary, the AdaT chooses to keep most of the Transformer block frozen and insert new adapter blocks. During training, it only updates these blocks and corresponding layer normalizations.}
  \Description{}
  \label{fig:FTA_vs_ADT}
  \vspace{-0.1in}
\end{figure}

\begin{enumerate}[leftmargin=*]
\item The standard fine-tuning often involves updating the entire pre-trained model. Thereby, in scenarios where RS provides services for multiple vertical channels as described in Figure~\ref{fig:Multi_cate}, TransRec has to maintain a  copy of the fine-tuned model for every channel. This largely hinders parameter-sharing across domains and brings additional costs in model updates, maintenance, and storage.
\item The large foundation model could quickly overfit when fully fine-tuning it on a small-scale downstream dataset.  Fine-tuning the last (few)  layers provide an alternative solution. However, it requires many manual attempts to determine the number of layers to be tuned as
it highly depends on the pre-trained model and transfer learning task.  This is particularly difficult for TransRec models since they could have multiple item encoders~\cite{wang2022transrec,wu2021two,wu2021mm}.

\item  Fine-tuning the entire ME model could be very expensive and sometimes nearly impractical (e.g., with the billion-level parameter size) in terms of GPU memory. 
For example, if we use BERT-large (around 1.3G) as the item ME, it typically requires around 10G memory to perform full parameter fine-tuning, e.g., with the item (e.g., news) description sequence length of 50 and batch size of 1, by using only one item ME in TransRec. 
\end{enumerate}

These issues of standard fine-tuning motivate us to explore parameter-efficient transfer learning techniques for TransRec. Recent work \cite{wang2020k,houlsby2019parameter, Sung_2022_CVPR, pfeiffer2020mad, pfeiffer2020adapterhub,ge2023algrnet,karimi2021compacter,chen2022vision,chen2022revisiting} in NLP and CV suggests that by adding several plug-in networks, i.e., adapter blocks, to the Transformer-style backbones, one can achieve comparable results to full parameter fine-tuning by only optimizing these adapters. Figure \ref{fig:FTA_vs_ADT} demonstrates the difference between fine-tuning all parameters (FTA) and adapter tuning (AdaT). Since the number of parameters in these adapters is extremely small compared with the backbone model, they can thereby achieve parameter-efficient transfer learning. For example, by applying the classic Houlsby adapter \cite{houlsby2019parameter}, the number of trainable parameters can be reduced to less than 3\% of the entire backbone network. 
Another advantage of the adapter-based approach is that it enables modular design and easily decouples task-specific parameters from the large backbone network. This mitigates the difficulties of the model maintenance and inconsistent update issues mentioned above. In addition, it can introduce more robustness and achieve improved stability effects during transfer learning
as indicated in~\cite{han2021robust}.

Nevertheless, in the recommender system fields, little work has investigated the adapter techniques. A closely related work is PeterRec \cite{yuan2020parameter}.  
However, PeterRec adopts IDRec as the backbone, where the largest amount of parameters is in the ID embedding layer rather than these middle layers in which the adapters are usually inserted.
 In fact, so far, it remains completely unknown \textit{whether adapter-based transfer learning is well-suited to the TransRec models when learning item representations from the raw modality features}. To this end, we ask the following sub-questions: 
 \begin{enumerate}[leftmargin=*]
    \item \textbf{Q(i) Does the Adapter-based TransRec perform comparably to typical fine-tuning based TransRec? Does this hold for items with different modalities?}
    To answer it, we conduct a rigorous comparative study for the adapter-based and fine-tuning based TransRec on two item modalities (i.e., texts and images) with two popular recommendation architectures (i.e., SASRec~\cite{kang2018self} and CPC~\cite{wang2022transrec,oord2018representation}) and four powerful ME (i.e., BERT, RoBERTa~\cite{liu2019roberta}, ViT and MAE~\cite{he2022masked}).
    \item \textbf{Q(ii) If Q(i) is true or partially true, what about the performance of these cleverly designed adapters developed in other communities for TransRec problems? }
    To answer this, we benchmark four adapters widely adopted in the NLP and CV literature. We also add the results of lora, prompt tuning, and layer-normalization tuning for a comprehensive comparison. 
    \item \textbf{Q(iii) Are there any factors that affect the performance of these adapter-based TransRec models?
    } 
    We report performance comparisons with different strategies regarding how and where to insert the adapters and whether to tune the corresponding normalization layers. 
\end{enumerate}
At last, we look at the data scaling effect of TransRec in the source and target domains to examine whether adapter tuning is beneficial when pre-training TransRec with larger datasets.

\begin{figure*}
  \centering
   \includegraphics[width=0.93\textwidth]{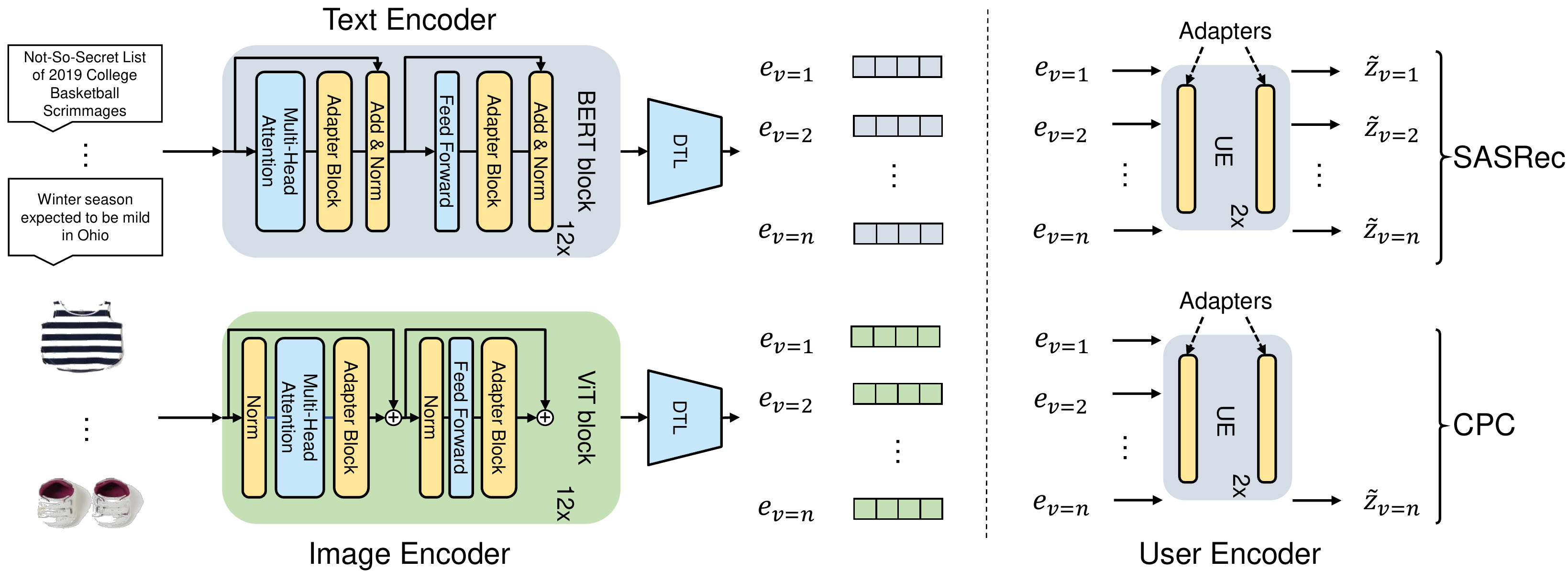}
  \caption{The adapter-based TransRec framework. The TransRec consists of a user encoder (UE) and multiple item encoders divided by the dotted line.  BERT and ViT are applied as examples of the text encoder and image encoder respectively. SASRec and CPC (DSSM variant) are used to train UE. $\tilde{Z}_{v=1},...,\tilde{Z}_{v=n}$ are vector generated by UE, $e_{v=1},...,e_{v={n}}$ are vectors generated by ME.
  Thereby, the way to inject adapters in UE follows the same way as that of the item encoder. 
  }
    \label{fig:Adapter_for_all} 
\end{figure*}

\section{Preliminary}
\label{sec:method}

\subsection{Overview of TransRec}
Given a recommendation dataset $\mathcal {D} = \{\mathcal {U}, \mathcal {V}, \mathcal {I}\}$ where $\mathcal {U}$, $\mathcal {V}$, $\mathcal {I}$ denote the set of users, the set of items and the set of interaction sequences,  respectively.  Like the typical recommendation task, we aim to predict the next item to be interacted by $u \in \mathcal {U}$ by exploiting his/her past behaviors $I_u$.  In TransRec, each item $v \in \mathcal {V}$ is associated with its raw modality features $\boldsymbol{m_v}$. By feeding $\boldsymbol{m_v}$ into an item ME $E_{item}$ (e.g., BERT for text or ViT for images), we obtain the vector representation of item $v$: 
\begin{equation}
\boldsymbol{z_v} = E_{item}(\boldsymbol{m_v})
\end{equation}

A basic dimension transformation Layer $(DTL)$ is added to ensure the consistency of the output dimensions of the item ME $E_{item}$ and input dimensions of the user encoder $E_{user}$: 
\begin{equation}
\boldsymbol{e_v} = DTL(\boldsymbol{z_v})
\end{equation}

Then, the representation of $u$ can be obtained through the user encoder $E_{user}$ (e.g., a Transformer
 backbone), which takes the interaction sequence $I_u$ as input: 
\begin{equation}
\widetilde{\boldsymbol{z_u}} =E_{user}\left((\boldsymbol{e_1}, \boldsymbol{e_2}, \cdots, \boldsymbol{e_{|I_u|}})\right)
\end{equation}

Finally, the next item to be interacted by user $u$ can be retrieved from $\mathcal {V}$ by the dot-product similarity between $\widetilde{\boldsymbol{z_u}}$ and ${\{\boldsymbol{z_v}\}}_{v=1}^{|\mathcal {V}|}$.

As mentioned, we aim to study the transfer learning problem 
by transferring the knowledge learned from the source domain $\mathcal {S}$ to the target domain $\mathcal {T}$. To be specific, a TransRec model is first trained on the source data $\mathcal {D}_s$ and then adapted to the target domain $\mathcal {T}$, usually by fine-tuning models using target data $\mathcal {D}_t$. Note that $\mathcal {D}_s$ and $\mathcal {D}_t$ do not necessarily contain overlapped items \& users. In this paper, we focus on parameter-efficient transfer learning from $\mathcal {S}$ to $\mathcal {T}$ by injecting the task-specific adapters into $E_{item}$ and $E_{user}$.

\subsection{Adapters for TransRec}
\noindent \textbf{Adapters overview.}
Adapters are task-specific neural modules inserted into a pre-trained model.
\cite{houlsby2019parameter} proposed to use a bottleneck network with a few parameters to project the original features to a lower dimension and then project them back after applying a non-linearity. With a residual connection, it can be illustrated as:
\begin{equation}
    Adapter(\boldsymbol{y}) = fcUp(RELU(fcDown(\boldsymbol{y})))+\boldsymbol{y}
    \label{eq:adapter}
\end{equation}
where $fcUp$ and $fcDown$ represent fully-connected layers that project the input dimensions up and down, respectively. 

\noindent \textbf{Adopting adapters in TransRec.} 
The TransRec architecture contains two sub-modules, namely, the item encoder $E_{item}$ and user encoder $E_{user}$, both of which are based on the Transformer blocks.  The architecture of adapter-based TransRec is illustrated in Figure~\ref{fig:Adapter_for_all}. For $E_{item}$ with textual modality (e.g., BERT), we follow the insertion strategy in \cite{houlsby2019parameter}, where two adapter blocks are inserted into each Transformer block, with one after the multi-head self-attention layer and the other after the feedforward network (FFN) layer. For $E_{item}$ with visual modality (e.g., ViT), the network structure remains the same, except for the position of LayerNorm. The user encoder $E_{user}$ also uses the same Transformer\footnote{One might wonder whether other networks can be used as TransRec backbone. In fact, TransRec that learns recommendation models directly from item \textit{raw} modality features (vs. ID features~\cite{yuan2021one}, vs. pre-extracted fixed features~\cite{hou2022towards} from ME) is still at a very early stage. Several existing literature ~\cite{shin2021scaling,wang2022transrec,rajput2023recommender,zhang2023ninerec,li2023exploring} is all based on the Transformer-style backbone, the most well-known SOTA sequential encoder. In practice, the Transformer backbone can be easily replaced with other sequential networks.
Second, can the CTR (click-through rate prediction)  models be used as TransRec backbones? Unfortunately, the classical one-tower CTR  models (e.g., DeepFM~\cite{ma2018modeling} \& MMOE~\cite{ma2018modeling}) 
 cannot be directly used as a  pre-training backbone for TransRec since some domain-specific features are not transferable or easily decoupled when adapting to other datasets. Instead, the two-tower DSSM model~\cite{wu2021empowering} can often be used to pre-train TransRec, as shown below.} architecture, the only difference is that $E_{user}$ is unidirectional here. In addition to this, it adopts the same insertion method as the item encoder.

\noindent \textbf{Training objectives.} 

In TransRec, $E_{user}$ takes the interaction sequence of user $u$ (denoted as $I_u$, with length $n$) as input, and outputs the hidden vectors of corresponding input elements, i.e.: 

\begin{equation}
    (\widetilde{\boldsymbol{z_1}}, \widetilde{\boldsymbol{z_2}}, …, \widetilde{\boldsymbol{z_n}}) = E_{user} \Bigl( (\boldsymbol{e_1}, \boldsymbol{e_2}, \cdots, \boldsymbol{e_n}) \Bigl)
\end{equation}
We use the SASRec~\cite{oord2018representation} and CPC~\cite{oord2018representation,wang2022transrec} framework
to train TransRec.
In SASRec, $E_{user}$ is expected to predict the corresponding next item of all elements in $I_u$, whereas, in the CPC framework, we only aim to predict the (n+1)-th item given the entire sequence. Note SASRec, in general, outperforms CPC in terms of accuracy, but CPC is essentially a more flexible two-tower based DSSM method~\cite{wang2022transrec} that is able to incorporate various user and item features.
Following \cite{kang2018self,wang2022transrec}, we apply the binary cross entropy (BCE) loss for both recommendation frameworks:

 \begin{equation}
    \left\{
    \begin{aligned}
        - \sum\limits_{u \in U} \sum\limits_{t \in N} \left[\log \sigma(\widetilde{\boldsymbol{z_t^u}} \cdot \boldsymbol{e_{t+1}^u}) + \log (1 - \sigma(\widetilde{\boldsymbol{z_t^u}} \cdot \boldsymbol{e_j}))\right] & \text{SASRec }\\
        - \sum\limits_{u \in U} \left[\log \sigma(\widetilde{\boldsymbol{z_n^u}} \cdot \boldsymbol{e_{n+1}^u}) + \log (1 - \sigma(\widetilde{\boldsymbol{z_n^u}} \cdot \boldsymbol{e_j}))\right] & \text{\,\,\,\,CPC}
    \end{aligned}\right.
    \label{eq:UE}
\end{equation}

where $\boldsymbol{e_j}$ denotes the embedding of a randomly sampled negative item from $\mathcal{V}$ and $j \not\in I_u$. $N$ represents the set of $[1, 2, \cdots, n]$ (see Figure~\ref{fig:Adapter_for_all}).

\section{Experiment Setup}
\textbf{Datasets.}  
We evaluate adapter-based TransRec with two modalities. For items with textual features, we utilize the MIND \cite{wu2020mind} English news recommendation dataset as the source domain, and the Adressa \cite{gulla2017adressa}, a Norwegian news recommendation dataset as the target domain.\footnote{We translate the Adressa dataset from Norwegian to English using Google Translate and randomly sample 20,000 users to construct the dataset. } For the visual modality, the Amazon review dataset for clothing\&shoes recommendation\cite{he2016ups, mcauley2015image} is used as the target domain, and the H\&M\footnote{\url{https://www.kaggle.com/competitions/h-and-m-personalized-fashion-recommendations}} personalized fashion recommendation dataset is used as the source domain. We select the latest 20 clicked news to construct interaction sequences for text recommendation tasks. Due to the constraint of GPU memory, the sequence length for fashion recommendation is limited to 10. After the preprocessing, the details of the datasets are shown in Table~\ref{tab:dataset}.

\begin{table}
  \caption{Dataset Description}
  \label{tab:dataset}
  \begin{tabular}{clcccc}
    \hline
    \multirow{2}{*}{Dataset}&\multirow{2}{*}{Users}&\multirow{2}{*}{Items}&\multirow{2}{*}{Interaction}&\multirow{2}{*}{Content}&\multirow{2}{*}{Domain}\\
    &&&&&\\
    \hline
    MIND&630,235&79,707&10,928,010&Text&Source\\
    Adressa&20,000&3,149&280,656&Text&Target\\
    H\&M&500,000&86,733&6,500,000&Image&Source\\
    Amazon&21,153&14,348&128,808&Image&Target\\
  \hline
\end{tabular}
\end{table}
\noindent\textbf{Evaluations.}
 Following previous works \cite{he2017neural}, we adopt the leave-one-out strategy to split the datasets: the last item in the interaction sequence is used for evaluation, and the item before the last is used as validation while the rest are for training. The HR@10 (hit ratio)  and NDCG@10 (Normalized Discounted Cumulative Gain) \cite{yuan2020parameter} are used as the evaluation metrics. Without special mention, all results are for the testing set.
 Note that we rank the predicted item with all items in the item set~\cite{krichene2020sampled}.

 \noindent\textbf{Implementation Details.} 
 The "bert-base-uncased", "roberta-base", "vit-base-patch16-224", and "vit-mae-base" from the Huggingface \footnote{https://huggingface.co/} platform are used as the text and image encoders, respectively. The dimension of hidden representations of the user encoder is searched in \{32,64,128\} and set to 64. The number of Transformer blocks and attention heads is fixed to 2. We apply Adam as the optimizer without weight decay throughout the experiments and extensively search the learning rate from 1e-6 to 1e-2 while keeping the dropout probability at 0.1.
 We set
 the batch size to 64 for textual datasets and 32 for visual datasets due to the GPU memory limits. 
 When adapting to the target domain, we set the batch size to 32 for both modalities.
 The hidden dimension of the adapter networks are carefully searched in \{8, 16, 32, 48, 64, 96, 128, 192, 384, 768\}, and the number of tokens of prompt tuning in \{5, 10, 20, 30, 40, 50\}.
 Note that the hyper-parameters of parameter-efficient modules are only searched in the SASRec-based architectures and directly transferred to the CPC-based methods.
 All hyper-parameters are determined according to the performance in the validation data. All results are reported on the testing set.

\section{Effectiveness of adapters in TransRec (Q(i))}
\label{sec:compare}
In this section, we evaluate the effectiveness of adapter tuning (AdaT) in TransRec since it is unknown whether AdaT works or not for recommendation models. 
Specifically, we run experiments on eight combinations: \{SASRec+BERT, CPC+BERT, SASRec+RoBERTa, CPC+RoBERTa, SASRec+ViT, CPC+ViT, SASRec+MAE, CPC+MAE\}, where BERT, RoBERTa, ViT, and MAE are
the most popular and widely accepted state-of-the-art (SOTA) ME in NLP and CV fields.

The most prevalent AdaT --- i.e., Houlsby \cite{houlsby2019parameter} Adapter --- results are present in Table \ref{tab:AdaT_vs_FTA}. Note that other adapter results are reported in the next benchmark section.  As can be seen,
TransRec, with the SASRec objective, consistently outperforms its CPC version. This is perhaps because SASRec is more powerful at modeling the item transition pattern in the user sequence and can thus alleviate the insufficient training data issue. For the text recommendation task, AdaT yields comparable results to fine-tuning all parameters (FTA) across evaluated frameworks (SASRec/CPC+BERT/RoBERTa) with a parameter reduction rate of over 97\%. However, for image recommendation, the performance gap between FTA and AdaT is relatively large, regardless of the training strategies used (SASRec/CPC+ViT/MAE). This result is somewhat justified, as the Houlsby adapter is primarily designed for NLP domain data and scenarios, which may make it suboptimal for visual tasks. 

\begin{figure}
  \centering
   \includegraphics[width=\linewidth]{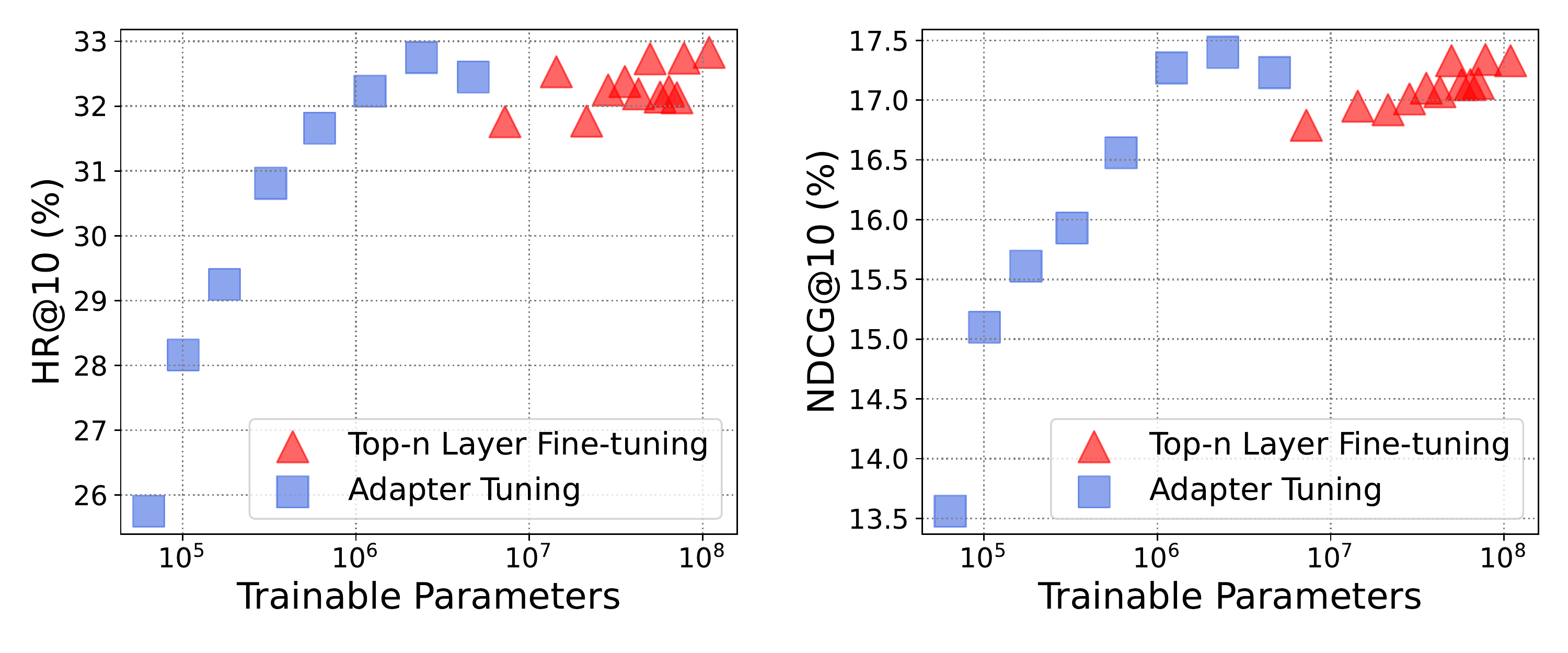}
  \caption{Adapter tuning vs. top layer tuning. Top layer tuning optimizes the top $k$ layers ($k=1,2, \cdots, 12$). Adapter tuning with an adapter size of $2^n$ ($n = 0, \cdots, 7$).
  }
  \vspace{-0.2in}
    \label{fig:top_vs_ada}
\end{figure}

To understand the impact of trainable parameters on domain adaptation, we simply test AdaT with different adapter sizes and compare its performance with top-n layer fine-tuning (FTN) for text recommendation. Since most of the trainable parameters come from item ME, we focus on the adapters in item ME and keep the UE with the original settings. The results are shown in Figure \ref{fig:top_vs_ada}, where the x-axis denotes the number of trainable parameters that are changed by gradually increasing/decreasing the hidden dimension of the adapter module (for AdaT) or by tuning more/fewer top Transformer layers (for FTN). Clearly, both FTN and AdaT  can improve performance with more trainable parameters. Furthermore, AdaT achieves competitive results with nearly two orders of magnitude fewer parameters than FTN.

\textbf{(Answer for Q(i)) Overall, when learning items with textual content,  TransRec with the SOTA AdaT realizes the parameter-efficient transfer from the source to the target domain, yielding comparable performance to FTA.
On the other hand, AdaT improves parameter efficiency at the cost of some performance drops for visual item recommendation but still could be an option in special cases where sufficient storage resources are unavailable. 
}
Therefore, how to design a specific adapter for various image-based recommendation models is a key research question.

\begin{table}[htbp]
\caption{Fine-tuning and adapter tuning comparison. FTA and AdaT represent "Fine-tune All" and "Adapter Tuning" respectively. TP stands for trainable parameters. All results of HR@10 and NDCG@10 in this table are denoted in the percentage (\%). \textbf{T} and \textbf{V} represent the textual and visual recommendation. The difference between FTA and AdaT is denoted by Diff.
}
\label{tab:AdaT_vs_FTA}
\begin{center}
\scalebox{1}{
\small
\begin{tabular}{c c | c  c | c}
\hline
\multirow{2}{*}{Architecture}&\multirow{2}{*}{Metrics}& \multirow{2}{*}{\,\,\,\,\,FTA\,\,\,\,\,} & \multirow{2}{*}{\,\,\,\,\,AdaT\,\,\,\,\,}& \multirow{2}{*}{Diff}\\
&&&\\
\hline
\multirow{2}{*}{SASRec+BERT(\textbf{T})}&HR@10 &\textbf{32.83} &32.52 &-0.94\%\\
&NDCG@10 &17.33  & \textbf{17.44}  &+0.63\%\\
\hline
\multirow{2}{*}{CPC+BERT(\textbf{T})}&HR@10 &29.56  &\textbf{30.07} &+1.69\%\\
&NDCG@10 &15.81 &\textbf{16.12} &+1.92\% \\
\hline
&TP&100\%&2.23\%&-97.77\%\\
\hline
\multirow{2}{*}{SASRec+RoBERTa(\textbf{T})}&HR@10 &32.02 &\textbf{33.14} &+3.38\%\\
&NDCG@10 &16.95  & \textbf{17.54}  &+3.36\%\\
\hline
\multirow{2}{*}{CPC+RoBERTa(\textbf{T})}&HR@10 &29.90  &\textbf{30.64} &+2.42\%\\
&NDCG@10 &15.86 &\textbf{16.20} &+2.10\% \\
\hline

&TP&100\%&1.95\%&-98.05\%\\
\hline
\hline
\multirow{2}{*}{SASRec+ViT(\textbf{V})}&HR@10 &\textbf{29.00} &27.66 &-4.59\% \\
&NDCG@10 &\textbf{25.61}  & 24.36  &-4.88\% \\
\hline
\multirow{2}{*}{CPC+ViT(\textbf{V})}&HR@10 &\textbf{26.56}  &25.29 &-4.78\% \\
&NDCG@10 &\textbf{22.09} &21.49 &-2.72\% \\
\hline
&TP&100\% & 2.82\% & -97.18\%\\
\hline
\multirow{2}{*}{SASRec+MAE(\textbf{V})}&HR@10 &\textbf{28.10}  &25.67 &-8.61\%\\
&NDCG@10 &\textbf{22.92} &21.99 &-4.05\% \\
\hline
\multirow{2}{*}{CPC+MAE(\textbf{V})}&HR@10 &\textbf{27.50}  &25.18 &-8.44\%\\
&NDCG@10 &\textbf{23.51} &21.83 &-7.14\% \\
\hline
&TP&100\% & 2.82\% &-97.18\%\\
\hline
\end{tabular}
}
\end{center}
\vspace{-0.3in}
\end{table}

\begin{table*}[htbp]
\caption{Benchmark popular parameter-efficient tuning techniques. Houlsby, K-Adapter, Pfeiffer, and Compacter adapters, along with LoRA, LayerNorm (LN), prompt tuning, are presented. The best approach to each architecture is marked in bold in this section. The “Architecture” is the combination of the user and item encoder. All results of HR@10 and NDCG@10 in this table are denoted in the percentage (\%). We represent the trainable parameters of each method by the percentage to the full fine-tuning. 
We omit the results of RoBERTa and MAE as ME, which are consistent with BERT \& ViT.
}
\label{tab:PEM_recommendation}
\begin{center}
\scalebox{1.0}{
\small
\begin{tabular}{c c | c  c c c c c c}
\hline
\multirow{2}{*}{Architecture}&\multirow{2}{*}{\,\,Metrics\,\,} & \multirow{2}{*}{\,\,Houlsby\,\,} &\multirow{2}{*}{\,\,Pfeiffer\,\,}&\multirow{2}{*}{\,\,Compacter\,\,}& \multirow{2}{*}{\,\,K-Adapter\,\,}& \multirow{2}{*}{\,LoRA\,}&\multirow{2}{*}{\,Prompt tuning\,}& \multirow{2}{*}{\,LN-tuning\,} \\
&&&&&&&\\
\hline
\multirow{2}{*}{SASRec+BERT}&HR@10 &\textbf{32.52}  &31.49&29.56&31.30& 30.89 &20.42&15.85\\
&NDCG@10  &\textbf{17.44}  &16.62 &15.32 &16.71 & 16.12 &11.02  &7.89  \\
\hline
\multirow{2}{*}{CPC+BERT}&HR@10 &\textbf{30.07} &29.32&26.40&28.49 & 27.60 &20.31 &15.81\\
&NDCG@10  &\textbf{16.12}  &15.37&13.73 &15.01 &14.52 &10.42 &8.65 \\
\hline
&Trainable Parameters&2.23\% &1.13\%&0.09\%&8.69\%&0.42\%&0.03\%&0.04\% \\
\hline
\multirow{2}{*}{SASRec+ViT}&HR@10 &\textbf{27.66}  &27.34&21.77 & 22.17&26.27&22.10 &19.89  \\
&NDCG@10  &\textbf{24.36}  &21.45&14.81 &13.65 &21.38 &15.72&13.49  \\
\hline
\multirow{2}{*}{CPC+ViT}&HR@10 &\textbf{25.29} & 24.63 &18.91  &18.05&22.29&19.43&18.79 \\
&NDCG@10 &\textbf{21.49}  &21.37&11.86&11.36 &10.71 & 13.99&13.00\\
\hline
&Trainable Parameters&2.82\%&1.43\%&0.12\%&11.03\% &0.54\% &0.07\%&0.05\% \\
\hline
\end{tabular}
}

\end{center}
\end{table*}

\begin{table}[htbp]
\caption{Comparison of full adapter-based TransRec and only adding adapters to the item or user encoder. Adapter$_{E_{i}}$ and Adapter$_{E_{u}}$ denote only adding adapters to the item and user encoder respectively. TP stands for trainable parameters. The subscripts $_V$ and $_B$ represent ViT and BERT. }
\vspace{-0.1in}
\label{tab:Ablation}
\begin{center}
\scalebox{1}{
\small
\begin{tabular}{c c | c | c  c}
\hline
\multirow{2}{*}{Architecture}&\multirow{2}{*}{Metrics} &\multirow{2}{*}{\,\,\,Adapter\,\,\,} & \multirow{2}{*}{\,Adapter$_{E_{i}}$\,} & \multirow{2}{*}{\,Adapter$_{E_{u}}$\,}\\
&&&&\\
\hline
\multirow{2}{*}{SASRec$_B$} & HR@10 &32.75 &32.45 &3.17\\
&NDCG@10  &17.40 &17.50  &1.58 \\
\hline
\multirow{2}{*}{CPC$_B$} & HR@10 &30.07  &29.56 &17.49\\
&NDCG@10 &16.12 &16.02  &9.34\\
\hline
\multirow{2}{*}{SASRec$_V$}&HR@10 &27.89 &15.79 & 5.78\\
&NDCG@10  &24.67 &10.51  &1.75 \\
\hline
\multirow{2}{*}{CPC$_V$}&HR@10 &25.48  &23.37 &9.57\\
&NDCG@10 &21.73  &19.32 &6.57 \\
\hline
&TP&100\%&99.64\%&0.36\%\\
\hline
\end{tabular}
}
\end{center}
\vspace{-0.2in}
\end{table}

\begin{table}
  \caption{Adapter position impact inside a Transformer block. We present the HR@10 for text and image recommendation with SASRec+BERT and SASRec+ViT architectures. Adapter$_{FFN}$ and Adapter$_{MHA}$ represent inserting the adapter block after FFN and MHA respectively. And Adapter$_{MHA}$++ and Adapter$_{FFN}$++ stand for the same architectures as the previous two but with 2x the parameters.}
  \label{tab:ada_pos_trm}
  \begin{tabular}{c | c  c c}
    \hline
    \multirow{2}{*}{Method}&\multirow{2}{*}{\,\,\,\,Text\,\,\,\,}&\multirow{2}{*}{\,\,Image\,\,}&\multirow{2}{*}{TP}\\
    &&&\\
    \hline
    Adapter$_{MHA+FFN}$ (Houlsby)&32.52&27.66&2,426,816\\
    Adapter$_{MHA}$ (Pfeiffer)&31.49&27.34&1,232,928\\
    Adapter$_{FFN}$&31.72&27.01&1,232,928\\
    Adapter$_{MHA}$++&31.71&27.49&2,417,472\\
    Adapter$_{FFN}$++&31.58&27.05&2,417,472\\
  \hline
\end{tabular}
\vspace{-0.15in}
\end{table}
\begin{table}[htbp]
\caption{Performance of the adapter insertion methods. The best HR@10 are marked in bold in each column. "w/" and "w/o" denote with and without updating the LayerNorm.}
\vspace{-0.1in}
\label{tab:P_AIM}
\begin{center}
\scalebox{1}{
\small
\begin{tabular}{c c |c  c |c c}
\hline

\multirow{2}{*}{Methods} & \multirow{2}{*}{LayerNorm} & \multirow{2}{*}{\,SASRec$_B$\,} & \multirow{2}{*}{\,CPC$_B$\,}&\multirow{2}{*}{\,SASRec$_V$\,}&\multirow{2}{*}{\,CPC$_V$\,}\\
&&&&&\\
\hline
\multirow{2}{*}{Serial}&- w/ LN &32.52 &30.07 &27.66 &25.30 \\
&- w/o LN&\textbf{32.75} &29.62 &\textbf{27.89} &\textbf{25.48} \\
\hline
\multirow{2}{*}{Parallel}&- w/ LN &32.70 &\textbf{30.28} &26.63 &24.92 \\
&- w/o LN &31.88 &29.66 &27.39 &24.92 \\
\hline
\end{tabular}
}
\end{center}
\vspace{-0.1in}
\end{table}
\section{Benchmarking Parameter-efficient Methods (Q(ii))}
\label{sec:benchmark}
In this section, we go one step further and benchmark four popular adapters in NLP and CV literature for applications in recommender systems. To be specific, we choose the Houlsby adapter \cite{houlsby2019parameter}, the K-Adapter \cite{wang2020k}, the Pfeiffer Adapter \cite{pfeiffer2020mad}, and Compacter \cite{karimi2021compacter} for evaluation. For a comprehensive comparison, we also include the results using prompt tuning \cite{jia2022visual}, LayerNorm tuning \cite{chen2022vision}, \textcolor{black}{and LoRA (Low-rank Adaptation) \cite{hu2021lora}}. We report the results in Table \ref{tab:PEM_recommendation}.

The structure of Houlsby is illustrated in Figure \ref{fig:FTA_vs_ADT}. The Pfeiffer architecture only inserts one adapter block for each Transformer block, saving about half of the parameters of the Houlsby. We adopt the implementation of the Pfeiffer adapter in \cite{karimi2021compacter}. The Compacter is constructed upon low-rank optimization and parameterized hypercomplex multiplication layers. The K-Adapter adds the adapters of Transformer structures to the backbone model in parallel. We insert two adapters for the item and user encoder respectively after structure searching. 
We also evaluate a popular prompt tuning \cite{lester2021power} technique as the baseline method, where the newly inserted token embeddings are added to the word embedding layer in the BERT model.
For visual recommendation, VPT \cite{jia2022visual} is used as the prompt tuning method, which adds the new token patch in the positional patch embedding for the ViT model. VPT needs to update the task-specific head compared to prompt tuning for text. 
We update the $DTL$ module as the task-specific head following the original setup.

The Houlsby adapter, among all methods, yields the best results under all settings with less than 3\% of trainable parameters of full fine-tuning. Following Houlsby, Pfeiffer achieves close performance with only around half of the parameters. This is because their adapter architectures are similar. The key difference is Pfeiffer removes adapters after FFN. 
The reason why Pfeiffer performs relatively worse will be further discussed in Section \ref{sec:bag_of_tricks}. The conclusion is that the position of adapter blocks does affect the overall performance.

 Compacter, with a special focus on parameter compression with low-rank factorization methodology, exhibits a significant decrease in recommendation accuracy, especially in image-based tasks.
This is most likely due to the extremely low capacity of trainable modules in Compacter. \textcolor{black}{The same occurrence can be seen in LoRA.} Figure \ref{fig:top_vs_ada} also shows that reducing parameters by a large amount can lead to very bad results. 
Therefore, TP (trainable parameters) matters in a certain range in the recommendation task. 

One exception here is the K-Adapter, which adopts a Transformer layer within the adapter module and requires much more parameters to train than its counterparts. Surprisingly, the performance drops severely.  We conjecture that the Transformer architecture within the K-Adapter is not suitable for domain adaptation since it was originally designed for knowledge injection rather than parameter-efficient purposes.
K-Adapter does not inject its information into the pre-trained model. Instead, it only receives knowledge from pre-trained models. The information flow direction makes its working mechanism very different.

The last two columns in Table \ref{tab:PEM_recommendation} show the results for prompt tuning, LayerNorm tuning. Prompt tuning offers a flexible way to utilize a big pre-trained model in various downstream tasks, mainly in the NLP domain. However, it fails to give competitive results as the adapters in TransRec.
LayerNorm tuning, i.e., only updating the LayerNorm parameters during adaptation also suffers from severe performance degradation. 
These results again potentially imply that TP are important for recommendation, although many NLP tasks can be performed well even with much fewer TPs.

\textbf{(Answer for Q(ii)) Overall, the Houlsby adapter obtains the best results in TransRec under all experimental settings, while the Pfeiffer adapter achieves slightly worse performance with half the number of parameters. } In the domain of NLP, Compacter yields significantly better performance than the popular Houlsby and Pfeiffer adapters, even with an extremely small amount of trainable parameters \cite{karimi2021compacter}. However, it fails to achieve decent results in the modality-based recommendation task. The LayerNorm tuning routinely performs worse than the full fine-tuning and adapter techniques in both CV and NLP with an accuracy drop of about 10\% to 20\% ~\cite{he2022parameter,houlsby2019parameter}; however, it 
is only half as accurate as the best Houlsby method in the recommendation scenarios. \textbf{ One key finding is that the adapter's trainable parameter size, insertion positions, and information flow directions are all key factors for the recommendation task.}
\vspace{-0.16in}

\section{Analysis of more factors (Q(iii))}
\label{sec:bag_of_tricks}
Since existing adapters are mainly derived from the NLP literature,  a natural challenge is to effectively apply them in the recommendation scenario. Specifically, we ask: \textit{where and how to insert the adapters for TransRec?} Regarding the question of \textit{where}, we aim to check whether the two modules in TransRec, $E_{item}$ and $E_{user}$, are equally important for domain transfer, as this is specific for the recommendation task. Regarding the question of \textit{how}, we evaluate two insertion strategies (serial vs. parallel) of the adapter networks and explore the effect of LayerNorm in the recommendation task.   

We first evaluate the effect of adapters inserted into different modules in TransRec. There are three ways to implement adapters:  placing them into both user and item encoders
($AdaT_{all}$),  only into the item encoder ($AdaT_{item}$), and into the user encoder ($AdaT_{user}$) (all other parameters are fixed).
From Table \ref{tab:Ablation}, first, we can clearly see that $AdaT_{item}$ outperforms $AdaT_{user}$ by a large margin in all experimental settings. This indicates that the item encoder plays a more important role in
 the recommendation task and requires more re-adaptation on the new datasets. Second, $AdaT_{item}$ achieves comparable results as  $AdaT_{all}$ in textual RS, suggesting that the knowledge stored in $E_{user}$ can be largely re-used with the adapted $E_{item}$.
 However, there is still a significant gap between $AdaT_{item}$ and $AdaT_{all}$ for the visual task, indicating that the parameter adaptation of $E_{user}$ is also important. Besides, This again shows that the image-based visual recommendation is a more difficult task than the text recommendation.
 
From Section~\ref{sec:benchmark}, we know that the Pfeiffer adapter shows a performance gap from Houlsby. The only difference is that Houlsby inserts adapter blocks after both FFN and MHA, whereas Pfeiffer only inserts after MHA. We further test the setting of this adapter to verify the impact of the position of insertion. The results are in Table \ref{tab:ada_pos_trm}. Thereby, accuracy drops may come from two reasons: 1) no tunable adapter after FFN; 2) fewer TP because of removing one adapter. To verify 1), we changed the position of the adapter (i.e., inserting after FFN), which yielded the same results. To verify 2), we double the number of TP, which still performs similarly, indicating adapters should be inserted for both FFN and MHA for RS. 
\begin{figure*}
  \centering
   \includegraphics[width=0.86\textwidth]{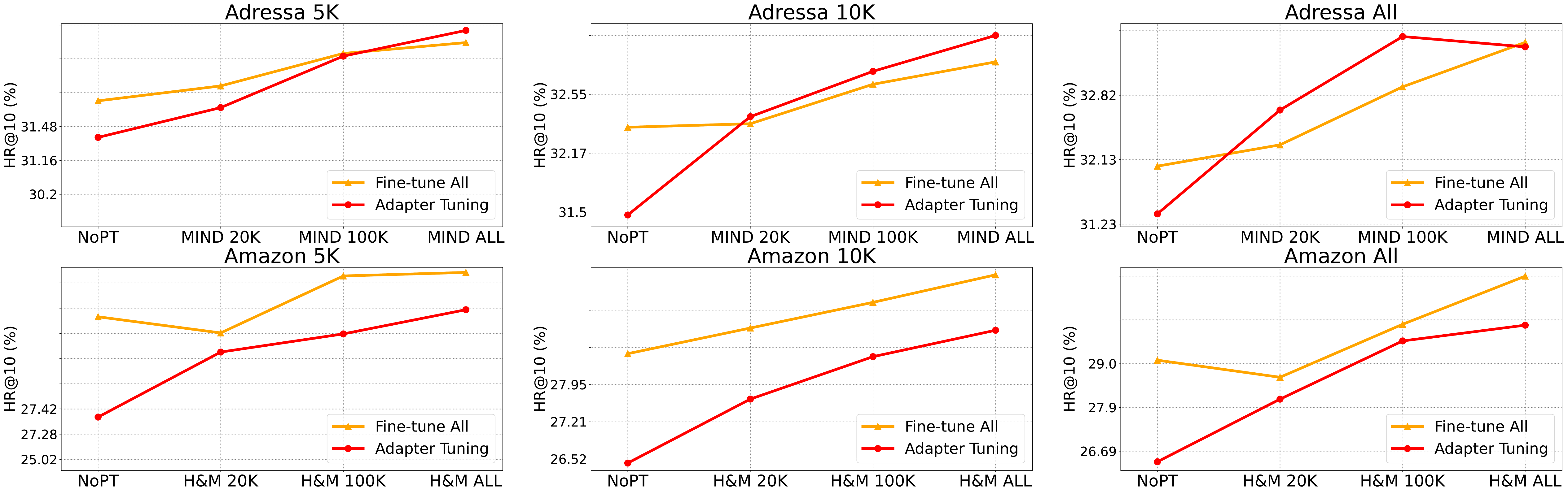}
\vspace{-0.1in}
  \caption{Scaling effects of fine-tuning and adapter-based TransRec using the SASRec objective. 
  The x-axis represents the size of pre-trained data. NoPT refers to TransRec that was not pre-trained by the source domain dataset.}
    \label{fig:scaling_plot}
\vspace{-0.15in}
\end{figure*}
We then compare two adaption insertion strategies: serial and parallel
The parallel approach is also adopted in~\cite{zhu2021serial,houlsby2019parameter}. In Table \ref{tab:P_AIM}, 
it can be seen that the two insertion methods, in general, perform very similarly. 
The other observation is that whether tuning the LayerNorm layer or not has almost no obvious influence on the recommendation accuracy. This is very different from other fields \cite{houlsby2019parameter, karimi2021compacter} where they strongly suggest optimizing both the adapter and LayerNorm layers for obtaining the optimal results. Thereby, for practical RS tasks, we only need to save the adapter modules, which is more efficient and convenient.

\textbf{(Answer for Q(iii)) We draw some conclusions here: (1)  TransRec
should place adapters for both user and item encoders for obtaining the optimal results, in particular for visual RS whose performance drop is very significant if the parameters of either the item encoder or user encoder are completely fixed; (2) the insertion position on the Transformer layers is  also important,  both FFN and MHA require a separate adapter module;}
(3) other factors such as insertion way (serial or parallel) and LayerNorm optimization do not matter a lot for the recommendation task, although they are often considered for NLP and CV tasks; (4) again, the number of trainable parameters is always a key factor for the accuracy of TransRec, certainly within a certain range, as described in the previous section.

\vspace{-0.1in}
\section{Scaling Effects}
\label{sec:scale}
To better understand the role of training data during pre-training and downstream adaption, we conduct experiments by scaling data in both the source domain ($\mathcal {D}_s$) and the target domain ($\mathcal {D}_t$), and present the results in Figure \ref{fig:scaling_plot}. According to the performance curves, we make the following observations: 

(1) Despite some exceptions, the HR@10 shows a clear trend of improvement for FTA and AdaT on the two modalities as the upstream pre-training dataset in  $\mathcal {D}_s$ increases. This observation has important implications: \textbf{for industrial recommender systems,}  \textbf{one can expect greater performance gains with more pre-trained source domain data.} 

(2) AdaT shows poor results under the NoPT setting, where only the item ME is pre-trained (on some NLP and CV data, e.g., ImageNet), and the user encoder is randomly initialized.
This explains that the lightweight adapter network indeed (or can only) does some parameter adaption work. It should fail or perform worse when the parameters (in the user encoder) are randomly initialized.

(3) There are some other observations consistent with the previous description. For example,
AdaT achieves comparable results to FTA for the text-based recommendation, while it lags behind FTA for image-based recommendation regardless of the size of the source and target datasets. We omit such repetitive descriptions here.

Imagine a practical scenario where we have a large number of user-item interactions in some industrial platform. The pre-trained knowledge (i.e., parameters) of TransRec in this platform can be effectively transferred to serve many other recommendation systems or channels through adapter tuning.

\vspace{-0.1in}
\label{ap:source}
\section{Related Work}
\noindent \textbf{Parameter-efficient transfer learning (PETL).} 
Researchers have been working on PETL for years to alleviate the gigantic amount of trainable parameters in large-scale pre-trained models. The principal way is to introduce adapter tuning techniques \cite{qin2022external,ma2022scattered}. In NLP, the first
adapter was proposed in~\cite{houlsby2019parameter} where authors uncovered that only training the newly inserted adapter blocks without any modification of the pre-trained parameters  could achieve competitive results to  full parameter fine-tuning. 
Pfeiffer et al. \cite{pfeiffer2020adapterhub} proposed the AdapterHub framework to facilitate, simplify, and speed
up transfer learning  across a variety of languages and tasks.
 \cite{wang2020k} proposed to inject multiple kinds of knowledge into large pre-trained models by K-Adapter. \textcolor{black}{Recently, \cite{zhang2023llama, gao2023llama} proposed LLaMA-Adapter for LLM (Large Language Models).}
He et al. \cite{he2022parameter} explored the PETL techniques in ViT-based computer vision tasks. ViT-Adapter~\cite{chen2022vision} achieved state-of-the-art performance on many dense prediction tasks of CV. \textcolor{black}{LoRA is another PETL technique similar to adapters but avoiding the inference latency \cite{hu2021lora}. Concurrent work \cite{geng2023vip5,bao2023tallrec} used adapter/LoRA to adapt LLM for textual RS, unlike our TranRec problem that transfers from a source to a target domain.}

Prompt \cite{zhang2023prompt,wang2022towards,ma2022xprompt, xin2022rethinking,wu2022personalized,li2022personalized,lester2021power,li2023prompt} is another popular PETL paradigm. It shows that only optimizing the embeddings of a few prompt tokens exhibits similar performance as the full model fine-tuning. Recently, P5~\cite{geng2022recommendation} presented a ``pretrain, personalized prompt \& predict paradigm'' that can learn multiple recommendation-related tasks together by formulating them as prompt-based natural language tasks. M6-Rec~\cite{cui2022m6} showed that prompt tuning outperformed fine-tuning with negligible  1\% task-specific parameters. However, in this paper, with the two popular architectures,  we found that the standard prompt tuning is still unsatisfactory compared to adapter- or fine- tuning.  \cite{wu2022selective} and \cite{wu2022personalized}  utilized prompt tuning to study the selective fairness and cold-start  recommendation, but are ID-based methods different from our modality setting. 

\noindent \textbf{Modality-based TransRec.} Inspired by the success of foundation models in NLP and CV fields, the modality-based/only recommendation (MoRec) has attracted rising attention recently \cite{wu2021mm,sun2020multi,yuan2023go,wang2022transrec,li2023exploring,zhang2023ninerec,wei2023multi,wang2023missrec,cheng2023image,ni2023content,liu2023id,qu2023thoroughly}. Typically, they use the foundation model, such as BERT, RoBERTa, and GPT \cite{brown2020language} as the text encoder or ViT, and ResNet as the image encoder. The user encoders still keep a similar fashion as the traditional IDRec architectures, e.g., SASRec \cite{kang2018self}, BERT4Rec~\cite{sun2019bert4rec}. 

A key advantage of MoRec models is that they are naturally transferable  because item modality representation is universal regardless of platforms and systems. For example,
ZESRec \cite{ding2021zero} proposed a zero-shot predictor  by leveraging the natural language representation extracted from BERT. Similar work also includes TransRec~\cite{wang2022transrec} by Wang et al, UniSRec~\cite{hou2022towards}, and ShopperBERT~\cite{shin2021one4all}, which all leveraged textual features to realize transferable recommendation.
However, so far,  existing TransRec literature (especially image TransRec) mostly utilizes the off-the-shelf features pre-extracted from ME, which has efficiency advantages over fine-tuning heavy ME. 
Recently, \cite{wang2022transrec,yuan2023go,li2023exploring}  started to perform joint training of user encoder and item ME in both pre-training and fine-tuning stages (i.e., our FTA baseline), which showed significantly improved performance compared to the pre-extracted fixed features. Therefore, in this paper, we study TransRec as a comparison baseline in a more powerful end-to-end (or joint) learning manner.\footnote{Some recent work also performed end-to-end training for modality recommendation, but they~\cite{wu2021empowering,xiao2022training,wu2022end} did not study the TransRec problem. We do not discuss them here.}

To the best of our knowledge, few studies have investigated the adapter tuning techniques for modality-based TransRec, especially for inserting adapters into item ME. PeterRec \cite{yuan2020parameter} proposed the first adapter tuning technique for the recommendation task, but it highly relies on overlapped userIDs when performing transfer learning. Moreover, the majority of parameters in PeterRec are on the ID embedding layer rather than the middle layers. Therefore, adapter tuning is still new for modality-based TransRec models.

\vspace{-0.1in}
\section{Conclusion and future work}
In this paper, we conducted an extensive empirical study examining the performance of the popular Adapter Tuning (AdaT) techniques for modality-based TransRec models.
We identified two facts: (1) the SOTA AdaT achieves competitive results compared to fine-tuning all parameters (FTA) for text recommendation; (2)  AdaT works fine but lags slightly behind FTA for image recommendations. 
We then benchmarked four well-known AdaT approaches and found that their behavior was somewhat idiosyncratic,  compared to NLP and CV tasks.  We deeply studied several key factors that may influence AdaT results for recommendation tasks.
At last, we found that TransRec with AdaT meets our expectations due to the ideal data scaling effect ---
 TransRec benefits when upscaling the source domain data or downscaling the target domain data. Our work provides  important guidelines for parameter-efficient
transfer learning for modality recommendation models. It also has important practical implications for foundation  models~\cite{bommasani2021opportunities} in the RS community, with the grand goal of `one model for all'~\cite{yuan2021one,sheng2021one,yuan2020parameter,wang2022transrec}.

There are several interesting future directions. The first one is to develop more advanced AdaT TransRec for visual item recommendation. Then, we are also interested in investigating the effects of AdaT for multimodal (i.e., both text and image) TransRec. Third, given that  most typical AdaT does not help to speed up the training process in practice (nor for NLP and CV tasks), it is important to explore effective optimization techniques to reduce the computational cost and time for TransRec through end-to-end training of item modality encoders.  

\bibliographystyle{ACM-Reference-Format}
\balance
\bibliography{sample-base}

\appendix
\clearpage

\end{document}